\documentclass[aps,a4paper,10pt,twocolumn,showpacs,showkeys]{revtex4}
\usepackage{graphicx}
\usepackage{amssymb}

\usepackage{graphicx}
\usepackage{amssymb}
\usepackage{amsfonts}
\usepackage[dvips]{color}

\newcommand{\ket}[1]{\vert #1 \rangle}
\newcommand{\bra}[1]{\langle #1 \vert}

\newcommand{\calpha}{\overline{\alpha}}
\newcommand{\cbeta}{\overline{\beta}}
\newcommand{\cxi}{\overline{\xi}}
\newcommand{\czeta}{\overline{\zeta}}
\newcommand{\cv}{\overline{v}}
\newcommand{\cw}{\overline{w}}
\begin{document}
\title{Degaussification of twin-beam and nonlocality in the phase space}
\date{\today}
\author{Stefano Olivares}
\email{Stefano.Olivares@mi.infn.it}
\author{Matteo G. A. Paris}
\email{Matteo.Paris@fisica.unimi.it} \affiliation{Dipartimento di
Fisica dell'Universit\`a degli Studi di Milano, Italy}
\begin{abstract}
We show that inconclusive photon subtraction (IPS) on twin-beam produces
non-Gaussian states that violate Bell's inequality in the phase-space. The
violation is larger than for the twin-beam itself irrespective of the IPS
quantum efficiency. The explicit expression of IPS map is given both for the
density matrix and the Wigner function representations.
\end{abstract}
\maketitle
\section{Introduction}\label{s:intro}
The twin-beam state (TWB) of two modes of radiation can be expressed
in the photon number basis as
\begin{equation}\label{twb:fock}
|\Lambda \rangle\rangle_{ab} = \sqrt{1-\lambda^2}\,
\sum_{n=0}^{\infty}\,\lambda^n |n,n\rangle_{ab}\,,
\end{equation}
where $\lambda = \tanh (r)$, $r$ being the TWB squeezing parameter.
TWB is described by a Gaussian Wigner function
\begin{eqnarray}
W_{r}(\alpha,\beta) &=&
\frac{4}{\pi^2}\exp\{
-2 A (|\alpha|^2+|\beta|^2)\nonumber\\
&\mbox{}& \hspace{1.5cm} + 2 B (\alpha\beta + \calpha\cbeta)\}\,,
\label{twb:wig}
\end{eqnarray}
with $A \equiv A(r) = \cosh(2 r)$ and $B \equiv B(r) = \sinh (2 r)$.
Since (\ref{twb:wig}) is positive-definite, TWB are not suitable to
test nonlocality through homodyne detection.
Indeed, the Wigner function itself provides an explicit hidden variable
model for homodyne measurements \cite{bana,sanchez}. On the other hand,
it has been shown \cite{bana} that TWB exhibits a nonlocal
character for parity measurements. This is known as nonlocality in the
phase-space since Bell inequalities can be written in terms of the Wigner
function, which in turn describes correlations for the joint measurement of
displaced parity operators. Overall, the positivity or the negativity
of the Wigner function has a rather weak relation to the locality or the
nonlocality of quantum correlations.
\par
In Ref. \cite{ips:tele} we have suggested a conditional measurement
scheme on TWB leading to a non Gaussian entangled mixed state,
which improves fidelity in the
teleportation of coherent states. This process, called inconclusive photon
subtraction (IPS), is based on mixing each mode of the TWB with the vacuum
in a unbalanced beam splitter and then performing inconclusive photodetection
on both modes, {\em i.e.} revealing the reflected beams without discriminating
the number of the detected photons.
\par
A single mode version of the IPS, mapping squeezed light onto
non-Gaussian states, has been recently realized experimentally
\cite{grangier}. Moreover, IPS has been suggested as a feasible
method to modify TWB and test nonlocality using homodyne detection
\cite{sanchez}.
\par
In this paper we address IPS as a {\em degaussification} map for TWB,
give its explicit expression for the density matrix and the Wigner
function, and investigate the nonlocality of the resulting state
in the phase-space.
\par
The paper is structured as follows. In Section \ref{s:bana} we
review nonlocality in the phase-space, {\em i.e.} Wigner function
Bell's inequality based on measuring the displaced parity operator
on two modes of radiation. In Section \ref{s:degauss} we
illustrate the IPS process as a degaussification map and calculate
the Wigner function of the IPS state. The nonlocality of the IPS
state in the phase-space is then analyzed in Section
\ref{s:nonlocal}, whereas in Section \ref{s:homodyne} we discuss
nonlocality using homodyne detection, extending the analysis of
Ref. {\cite{sanchez}}. Section \ref{s:remarks} closes the paper
with some concluding remarks.
\section{Nonlocality in the phase-space}\label{s:bana}
The displaced parity operator on two modes is defined as
\begin{eqnarray}
\hat{\Pi}(\alpha,\beta) &=&
D_a(\alpha)(-1)^{a^\dag a}D_a^\dag(\alpha)\nonumber\\
&\mbox{}& \hspace{1.5cm} \otimes D_b(\beta)(-1)^{b^\dag b}D_b^\dag(\beta)\,,
\end{eqnarray}
where $\alpha, \beta \in {\mathbb C}$, $a$ and $b$ are mode
operators and $D_a(\alpha)=\exp\{\alpha a^\dag - \overline{\alpha}
a\}$ and $D_b(\beta)$ are single-mode displacement operators.
Parity is a dichotomic variable and thus can be used to establish
Bell-like inequalities \cite{CHSH}. Since the two-mode Wigner
function $W(\alpha,\beta)$ can be expressed as
\begin{equation}
W(\alpha,\beta) = \frac{4}{\pi^2}\, \Pi(\alpha,\beta)\,,
\end{equation}
$\Pi(\alpha,\beta)$ being the expectation value of
$\hat\Pi(\alpha,\beta)$, the violation of these inequalities is also
known as nonlocality in the phase-space. The quantity involved in such
inequalities can be written as follows
\begin{eqnarray}
\mathcal{B} &=& \Pi(\alpha_1,\beta_1)+ \Pi(\alpha_2,\beta_1)\nonumber\\
&\mbox{}& \hspace{.5cm}+ \Pi(\alpha_1,\beta_2)-
\Pi(\alpha_2,\beta_2)\,,
\label{bell:general}
\end{eqnarray}
which, for local theories, satisfies the condition $|\mathcal{B}|\le 2$.
\par
Following Ref. \cite{bana}, one can choose a particular set of
displaced parity operators, arriving at the following combination
\begin{eqnarray}
B(J) &=& \Pi(0,0)+ \Pi(\sqrt{J},0)\nonumber\\
&\mbox{}& \hspace{.5cm}+ \Pi(0,-\sqrt{J})-
\Pi(\sqrt{J},-\sqrt{J})\,,
\label{bell:bana}
\end{eqnarray}
which only depends on the positive parameter $J$,
characterizing the magnitude of the displacement.
If we evaluate the quantity (\ref{bell:bana}) in the case of the TWB, we
find that it exceeds the upper bound imposed by local theories
for a certain region of values $(J,r)$, its maximum being
$B \approx 2.19$ \cite{bana}.
\par
On the other hand, the choice of the parameters leading to Eq.
(\ref{bell:bana}) is not the best one, and the violation of the
inequality $|\mathcal{B}|\le 2$ can be enhanced using a different
parameterization \cite{ferraro}. A better result is achieved for
\begin{eqnarray}
C(J) &=& \Pi(\sqrt{J},-\sqrt{J})+ \Pi(-3\sqrt{J},-\sqrt{J})\nonumber\\
&\mbox{}& \hspace{.5cm}+ \Pi(\sqrt{J},3\sqrt{J})-
\Pi(-3\sqrt{J},3\sqrt{J})\,,
\label{bell:ale}
\end{eqnarray}
which, for the TWB, gives a maximum $C \approx 2.32$, greater than
the value $2.19$ obtained in Ref. \cite{bana}.
\par
In the following Sections we will see that the violation of the inequalities
$|B(J)|\le 2$ and $|C(J)|\le 2$  can be enhanced by degaussification of
the TWB.
\section{The degaussification process}\label{s:degauss}
The degaussification of a TWB can be achieved by subtracting
photons from both modes \cite{opatr,coch,ips:tele}. In Ref.
\cite{ips:tele} we referred to this process as to inconclusive
photon subtraction and showed that the resulting state, the IPS
state $\varrho_{\hbox{\tiny IPS}}$, can be used to enhance the
teleportation fidelity of coherent states for a wide range of the
experimental parameters.
\par
\begin{figure}
\includegraphics[scale=.7]{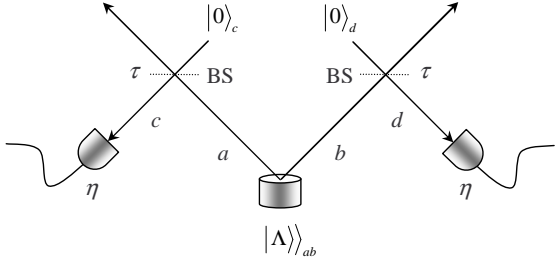}
\caption{\label{f:IPS:scheme} Scheme of the IPS process.}
\end{figure}
The IPS scheme is sketched in Fig. \ref{f:IPS:scheme}. The two
modes, $a$ and $b$, of the TWB are mixed with the vacuum (modes
$c$ and $d$, respectively) at two unbalanced beam splitters (BS)
with equal transmissivity $\tau = \cos^2\phi$; the  modes $c$ and
$d$ are then revealed by avalanche photodetectors (APD) with equal
efficiency $\eta$. APD's can only discriminate the presence of
radiation from the vacuum. The positive operator-valued measure
(POVM) $\{\Pi_0(\eta),\Pi_1(\eta)\}$ of each detector is given by
\begin{equation}\label{povm1}
  {\Pi}_0 (\eta) = \sum_{j=0}^{\infty} (1-\eta)^j
  \ket{j}\bra{j}\:,
\quad   {\Pi}_1(\eta) = \mathbb{I} - {\Pi}_0 (\eta)\:,
\end{equation}
$\eta$ being the quantum efficiency. Overall, the conditional measurement
on the modes $c$ and $d$, is described by the POVM (we are assuming the
same quantum efficiency for both photodetectors)
\begin{eqnarray}
{\Pi}_{00} (\eta)&=&{\Pi}_{0,c} (\eta) \otimes {\Pi}_{0,d}
(\eta)\:, \\
{\Pi}_{01} (\eta) &=& {\Pi}_{0,c} (\eta) \otimes
{\Pi}_{1,d} (\eta)\:, \\
{\Pi}_{10} (\eta)&=&{\Pi}_{1,c} (\eta) \otimes {\Pi}_{0,d}
(\eta)\:, \\
{\Pi}_{11} (\eta)&=&{\Pi}_{1,c} (\eta) \otimes
{\Pi}_{1,d} (\eta) \label{povm11}\;.
\end{eqnarray}
When the two photodetectors jointly click, the conditioned output state
of modes $a$ and $b$ is given by
\begin{widetext}
\begin{eqnarray}
\mathcal{E}(R)
= \frac{1}{p_{11}(r,\phi,\eta)}\hbox{Tr}_{cd}\big[
U_{ac}(\phi)\otimes U_{bd}(\phi) \: R
\otimes |0\rangle_c{}_{c}\langle 0|
\otimes |0\rangle_d{}_{d}\langle 0|
\: U_{ac}^{\dag}(\phi)\otimes U_{bd}^{\dag}(\phi) \:
{\mathbb I}_a \otimes
{\mathbb I}_b \otimes
{\Pi}_{11} (\eta)
\big]\:, \label{ptr}
\end{eqnarray}
where $U_{ac}(\phi)=\exp\{-\phi(a^{\dag} c-a c^{\dag}) \}$ and
$U_{bd}(\phi)$ are the evolution operators of the beam splitters
and $R$ the density operator of the two-mode state entering
the beam splitters (in our case $R = \varrho_{\hbox{\tiny TWB}} =
|\Lambda\rangle\rangle_{ab}{}_{ba} \langle\langle \Lambda |$). The partial
trace on modes $c$ and $d$ can be explicitly evaluated, thus
arriving at the Kraus decomposition of the IPS map.
We have
\begin{eqnarray}
\mathcal{E}(R)
= \frac{1}{p_{11}(r,\phi,\eta)}\:
\sum_{p,q=1}^{\infty}\:m_p(\phi,\eta)\:M_{pq}(\phi)\: R \:
M_{pq}^{\dag}(\phi)\: m_q(\phi,\eta)\:
\end{eqnarray}
\end{widetext}
with
\begin{equation}
m_p(\phi,\eta) =
{\displaystyle \frac{\tan^{2p}\phi\,\,[1-(1-\eta)^p]}{p!}}\,,
\end{equation}
and
\begin{equation}
M_{pq}(\phi) = \frac{\mbox{}}{\mbox{}}
a^p b^q \, (\cos\phi)^{a^\dag a + b^\dag b}\,,
\end{equation}
and
\begin{eqnarray}\label{dc:fock}
p_{11}(r,\phi,\eta) = {\rm Tr}_{ab}[ \mathcal{E}(R) ]
\end{eqnarray}
is the probability of a click in both detectors.
\par
Now, in order to investigate the nonlocality of the state
$\varrho_{\hbox{\tiny{IPS}}}=
\mathcal{E}(\varrho_{\hbox{\tiny TWB}})$ in the
phase-space, we explicitly calculate its Wigner function, which, as one
may expect, is no longer Gaussian and positive-definite.
\par
The state entering the two beam splitters is described by the Wigner
function
\begin{widetext}
\begin{equation}
W_{r}^{\hbox{\tiny (in)}}(\alpha,\beta,\zeta,\xi) =
W_{r}(\alpha,\beta)\,
\frac{4}{\pi^2} \exp\left\{ -2|\zeta|^2 - 2|\xi|^2 \right\}\,,
\end{equation}
where the second factor at the rhs represents the two vacuum states of
modes $c$ and $d$.
The action of the beam splitters on $W^{\hbox{\tiny (in)}}_{r}$ can be
summarized by the following change of variables
\begin{eqnarray}
\alpha &\to& \alpha\cos\phi + \zeta\sin\phi\,,\quad
\zeta  \to \zeta\cos\phi  - \alpha\sin\phi\,,\\
\beta  &\to& \beta\cos\phi  + \xi\sin\phi\,,\quad
\xi    \to \xi\cos\phi    - \beta\sin\phi\,,
\end{eqnarray}
and the output state, after the beam splitters, is then given by
\begin{eqnarray}
W_{r,\phi}^{\hbox{\tiny (out)}}(\alpha,\beta,\zeta,\xi) &=&
\frac{4}{\pi^2}\, W_{r,\phi}(\alpha,\beta)\,
\exp\left\{ -a |\xi|^2 + w \xi + \cw \cxi \right\}\nonumber\\
&\mbox{}& \hspace{.5cm}
\times\exp\left\{ -a |\zeta|^2 + (v + 2 B \xi \sin^2\phi)\zeta +
(\cv + 2 B \cxi \sin^2\phi)\czeta \right\}\,,
\end{eqnarray}
where
\begin{equation}
W_{r,\phi}(\alpha,\beta) =
\frac{4}{\pi^2}\,
\exp\left\{ -b (|\alpha|^2 + |\beta|^2)
+ 2 B \cos^2\phi\, (\alpha\beta + \calpha\cbeta) \right\}
\end{equation}
and
\begin{eqnarray}
a &\equiv& a(r,\phi) = 2 (A \sin^2\phi + \cos^2\phi),\\
b &\equiv& b(r,\phi) = 2 (A \cos^2\phi + \sin^2\phi)\,,\\
v &\equiv& v(r,\phi) = 2 \cos\phi\, \sin\phi\, [(1-A)\calpha + B \beta],\\
w &\equiv& w(r,\phi) = 2 \cos\phi\, \sin\phi\, [(1-A)\cbeta + B \alpha]\,.
\end{eqnarray}
\end{widetext}
\par
At this stage conditional on/off detection is performed on modes
$c$ and $d$ (see Fig. \ref{f:IPS:scheme}). We are interested in
the situation when both the detectors click. The Wigner function
of the double click element $\Pi_{11}(\eta)$ of the POVM (see Eq.
(\ref{povm11})) is given by \cite{ips:tele,cond:cola}
\begin{eqnarray}
W_{\eta}(\zeta,\xi) &\equiv& W[\Pi_{11}(\eta)](\zeta,\xi)\nonumber\\
&=& \frac{1}{\pi^2}\{
1-Q_{\eta}(\zeta)-Q_{\eta}(\xi)\nonumber\\
&\mbox{}&\hspace{1.5cm}
+Q_{\eta}(\zeta) Q_{\eta}(\xi)
\}\,,
\end{eqnarray}
with
\begin{equation}
Q_{\eta}(z) = \frac{2}{2-\eta}\,
\exp\Bigg\{-\frac{2\eta}{2-\eta}\, |z|^2 \Bigg\}\,.
\end{equation}
Using Eq. (\ref{ptr}) and the phase-space expression of trace
(for each mode)
\begin{equation}
{\rm Tr}[ O_1 O_2 ] = \pi\,
\int_{\mathbb{C}} d^2 z\, W[O_1](z)\, W[O_2](z)\,,
\end{equation}
$O_1$ and $O_2$ being two operators and $W[O_1](z)$ and $W[O_2](z)$ their
Wigner functions, respectively, the Wigner function of the output state,
conditioned to the double click event, is then given by
\begin{equation}\label{w:ips:informal}
W_{r,\phi,\eta}(\alpha,\beta) =
\frac{f_{r,\phi,\eta}(\alpha,\beta)}{p_{11}
(r,\phi,\eta)}\,,
\end{equation}
where
\begin{widetext}
\begin{equation}\label{w:ips:informal:f}
f_{r,\phi,\eta}(\alpha,\beta) =
\pi^2\,\int_{\mathbb{C}^2}d^2\zeta\,d^2\xi\,
\frac{4}{\pi^2}\,W_{r,\phi}(\alpha,\beta)\,
\sum_{j=1}^4 \frac{C_j(\eta)}{\pi^2}\,
G_{r,\phi,\eta}^{(j)}(\alpha,\beta,\zeta,\xi)\,,
\end{equation}
and $p_{11}(r,\phi,\eta)$ is the double-click probability (\ref{dc:fock}),
which can be written as function of $f_{r,\phi,\eta}(\alpha,\beta)$ as
follows
\begin{equation}\label{w:ips:informal:p}
p_{11}(r,\phi,\eta) =
\pi^2\,\int_{\mathbb{C}^2}d^2\alpha\,d^2\beta\,
f_{r,\phi,\eta}(\alpha,\beta)\,.
\end{equation}
The quantity $G_{r,\phi,\eta}^{(j)}(\alpha,\beta,\zeta,\xi)$
appearing in Eq. (\ref{w:ips:informal:f}) is
\begin{eqnarray}
G_{r,\phi,\eta}^{(j)}(\alpha,\beta,\zeta,\xi) &=&
\exp\left\{ -x_j |\zeta|^2 + (v + 2 B \xi \sin^2\phi)\zeta +
(\cv + 2 B \cxi \sin^2\phi)\czeta \right\}\nonumber\\
&\mbox{}& \hspace{1cm}
\times\exp\left\{ -y_j |\xi|^2 + w \xi + \cw \cxi \right\}\,,
\label{meas:int}
\end{eqnarray}
and the expressions of $C_j(\eta)$, $x_j\equiv x_j(r,\phi,\eta)$ and
$y_j\equiv y_j(r,\phi,\eta)$ are given in Table \ref{t:jj}.
\begin{table}[t!]
\caption{\label{t:jj}}
\begin{ruledtabular}
\begin{tabular}{cccc}
$j$ & $x_j(r,\phi,\eta)$ & $y_j(r,\phi,\eta)$ & $C_j(\eta)$
\vspace{.2cm}\\ \hline
1   &  $a$    & $a$ & 1 \\
2   &  $a + \frac{2}{2-\eta}$ & $a$ & $-\frac{2}{2-\eta}$ \\
3   &  $a$  & $a + \frac{2}{2-\eta}$ & $-\frac{2}{2-\eta}$ \\
4   &  $a + \frac{2}{2-\eta}$ & $a + \frac{2}{2-\eta}$
 & $(\frac{2}{2-\eta})^2$ \\
\end{tabular}
\end{ruledtabular}
\end{table}
\end{widetext}
\par
The mixing with the vacuum in a beam splitter with transmissivity $\tau$
followed by on/off detection with quantum efficiency $\eta$ is equivalent
to mixing with an effective transmissivity \cite{ips:tele}
\begin{equation}
\tau_{\rm eff} \equiv
\tau_{\rm eff}(\phi,\eta) = 1 - \eta (1-\tau)
\label{taueff}
\end{equation}
followed by an ideal ({\em i.e.} efficiency equal to 1) on/off
detection. Therefore, the state (\ref{w:ips:informal}) can be studied
for $\eta = 1$ and replacing $\tau$ with $\tau_{\rm eff}$.
Thanks to this substitution, after the integrations we have
\begin{widetext}
\begin{eqnarray}
f_{r,\phi,\eta}(\alpha,\beta) &=&
\frac{1}{\pi^2}\,
\sum_{j=1}^4 \frac{16 C_j(\eta)}{x_j y_j - 4 B^2 (1-\tau_{\rm eff})^2}
\nonumber\\
&\mbox{}&\hspace{.5cm}
\times\exp\{
-(b-f_j)|\alpha|^2 - (b-g_j)|\beta|^2
+(2 B \tau_{\rm eff} + h_j)(\alpha\beta + \calpha\cbeta)\}
\end{eqnarray}
and
\begin{equation}
p_{11}(r,\phi,\eta) =
\sum_{j=1}^4 \frac{16 [x_j y_j - 4 B^2 (1-\tau_{\rm eff})^2]^{-1} C_j(\eta)}{
(b-f_j)(b-g_j)-(2 B h_j \tau_{\rm eff})^2}
\,,
\end{equation}
where we defined
\begin{eqnarray}
f_j &\equiv& f_j(r,\phi,\eta) =
N_j
\, [x_j (1-A)^2 + 4 B^2 (1-A) (1-\tau_{\rm eff}) + y_j B^2]\,,\\
g_j &\equiv& g_j(r,\phi,\eta) =
N_j
\, [x_j B^2 + 4 B^2 (1-A) (1-\tau_{\rm eff}) + y_j (1-A)^2]\,,\\
h_j &\equiv& h_j(r,\phi,\eta) =
N_j
\, [(x_j + y_j) B (1-A) + 4 B (B^2 + (1-A)^2) (1-\tau_{\rm eff})]\,,\\
N_j &\equiv&  N_j(r,\phi,\eta) =
{\displaystyle
\frac{4 \tau_{\rm eff}\, (1-\tau_{\rm eff})}{x_j y_j - 4 B^2
(1-\tau_{\rm eff})^2}\,.
}
\end{eqnarray}
In this way, the Wigner function of the IPS state can be rewritten as
\begin{eqnarray}\label{ips:wigner}
W_{r,\phi,\eta}(\alpha,\beta) =
W_{r,\phi}(\alpha,\beta) \,
\sum_{j=1}^4\, \frac{4 C_j(\eta) K_{r,\phi,\eta}^{(j)}(\alpha,\beta)}
{p_{11}(r,\phi,\eta)\,[x_j y_j - 4 B^2 (1-\tau_{\rm eff})^2]}\,,
\end{eqnarray}
where we introduced
\begin{eqnarray}
K_{r,\phi,\eta}^{(j)}(\alpha,\beta) =
\exp\{ f_j |\alpha|^2 + g_j |\beta|^2
+ h_j (\alpha\beta + \calpha\cbeta)\}\,.
\end{eqnarray}
The state given in Eq. (\ref{ips:wigner}) is no longer a Gaussian state.
\end{widetext}
\section{Nonlocality of the IPS state}\label{s:nonlocal}
In this Section we investigate the nonlocality of the state
(\ref{ips:wigner}) in phase-space using the quantity $\mathcal{B}$
given in Eq. (\ref{bell:general}), referring to both the
parameterizations $B(J)$ (see Eq. (\ref{bell:bana})) and $C(J)$
(see Eq. (\ref{bell:ale})).
\par
\begin{figure}[t!]
\includegraphics[scale=.5]{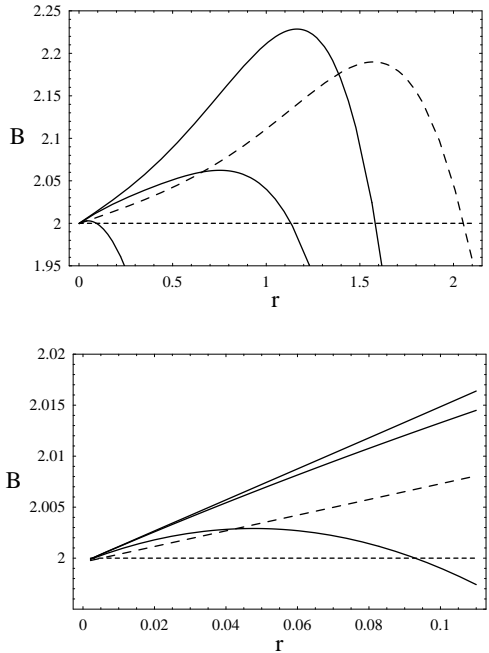}
\caption{\label{f:ips:vs:twb:j} Plot of $B(J)$ given in Eq.
(\ref{bell:bana}) for $J = 10^{-2}$. The dashed line is
$B_{r}^{\hbox{\tiny (TWB)}}(J)$,
while the solid lines are $B_{r,\phi,\eta}^{\hbox{\tiny (IPS)}}(J)$
for different values of $\tau_{\rm eff}$ (see the text):
from top to bottom $\tau_{\rm eff} = 0.999, 0.99$ and
$0.9$. When $\tau_{\rm eff} = .999$, the maximum of
$B^{\hbox{\tiny (IPS)}}_{r,\phi,\eta}(J)$ is $2.23$. The lower plot is a
magnification of the region $0\le r \le 0.11$ of the upper one.
Notice that for small $r$ there is always a region where
$B_{r}^{\hbox{\tiny (TWB)}}(J) < B_{r,\phi,\eta}^{\hbox{\tiny (IPS)}}(J)$.}
\end{figure}
\begin{figure}[t!]
\includegraphics[scale=.5]{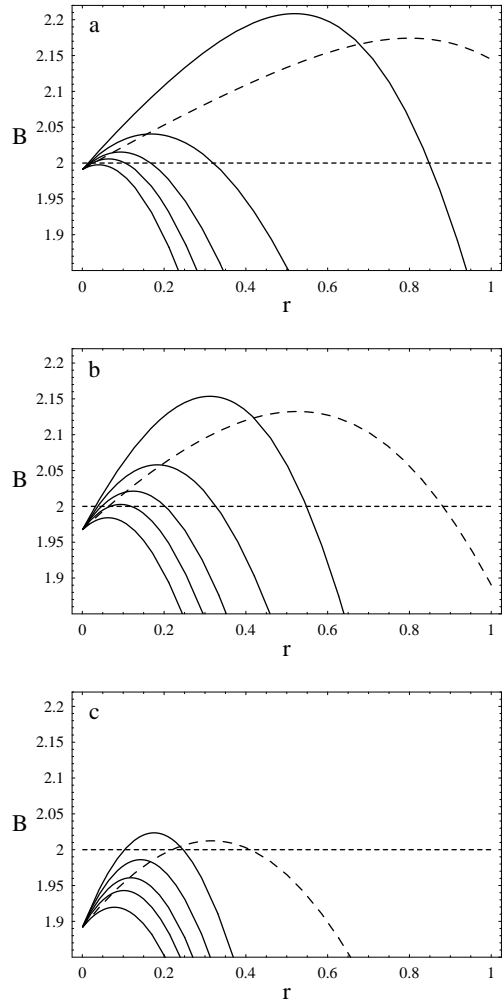}
\caption{\label{f:ips:vs:twb} Plots of $B(J)$ given in Eq.
(\ref{bell:bana}) as a function of
the squeezing parameter $r$ for different values of $J$:
(a) $J=5\,10^{-2}$, (b) $J=10^{-1}$ and (c) $J=2\,10^{-1}$.
In all the plots the dashed line is $B_{r}^{\hbox{\tiny (TWB)}}(J)$,
while the solid lines are $B_{r,\phi,\eta}^{\hbox{\tiny (IPS)}}(J)$
for different
values of $\tau_{\rm eff}$ (see the text): from top to bottom
$\tau_{\rm eff} = 0.999, 0.9, 0.8, 0.7$ and $0.5$.
Notice that there is always  a region for small $r$ where
$B_{r}^{\hbox{\tiny (TWB)}}(J) < B_{r,\phi,\eta}^{\hbox{\tiny (IPS)}}(J)$.
When $\tau_{\rm eff} = 0.999$ the maximum of
$B_{r,\phi,\eta}^{\hbox{\tiny (IPS)}}(J)$
is always greater than the one of $B_{r}^{\hbox{\tiny (TWB)}}(J)$.}
\end{figure}
As for a TWB, the violation of the Bell's inequality is observed
for small $r$ \cite{bana}. From now on, we will refer to $B(J)$ as
$B_{r}^{\hbox{\tiny (TWB)}}(J)$ when it is evaluated for a TWB
(\ref{twb:wig}), and as $B_{r,\phi,\eta}^{\hbox{\tiny (IPS)}}(J)$
when we consider the IPS state (\ref{ips:wigner}). We plot
$B_{r}^{\hbox{\tiny (TWB)}}(J)$ and $B_{r,\phi,\eta}^{\hbox{\tiny
(IPS)}}(J)$ in the Figs. \ref{f:ips:vs:twb:j} and
\ref{f:ips:vs:twb} for different values of the effective
transmissivity $\tau_{\rm eff}$ and of the parameter $J$: for not
too big values of the squeezing parameter $r$, one has that
$2<B_{r,\phi,\eta}^{\hbox{\tiny (TWB)}}(J)<
B_{r,\phi,\eta}^{\hbox{\tiny (IPS)}}(J)$. Moreover, when
$\tau_{\rm eff}$ approaches unit, {\em i.e.} when at most one
photon is subtracted from each mode, the maximum of
$B_{r,\phi,\eta}^{\hbox{\tiny (IPS)}}$ is always greater than the
one obtained using a TWB. A numerical analysis shows that in the
limit $\tau_{\rm eff} \to 1$ the maximum is $2.27$, that is
greater than the value $2.19$ obtained for a TWB \cite{bana}. The
limit $\tau_{\rm eff} \to 1$ corresponds to the case of one single
photon subtracted from each mode \cite{opatr,coch}. Notice that
increasing $J$ reduces the interval of the values of $r$ for which
one has the violation. For large $r$ the best result is thus
obtained with the TWB since, as the energy grows, more photons are
subtracted from the initial state \cite{ips:tele}. Since the
relevant parameter for violation of Bell inequalities is
$\tau_{\rm eff}$, we have, from Eq. (\ref{taueff}), that the IPS
state is nonlocal also for low quantum efficiency of the IPS
detector.
\par
\begin{figure}[t!]
\includegraphics[scale=.5]{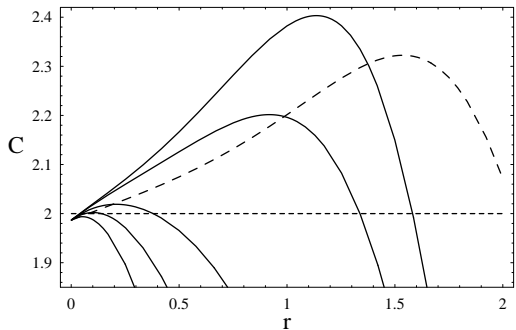}
\caption{\label{f:ips:ale} Plots of $C(J)$ given in Eq.
(\ref{bell:ale}) as a function of
the squeezing parameter $r$ for $J = 1.6\, 10^{-4}$.
In all the plots the dashed line is $C_{r}^{\hbox{\tiny (TWB)}}(J)$,
while the solid lines are $C_{r,\phi,\eta}^{\hbox{\tiny (IPS)}}(J)$
for different
values of $\tau_{\rm eff}$ (see the text): from top to bottom
$\tau_{\rm eff} = 0.999, 0.99, 0.95, 0.9$ and $0.8$.
When $\tau_{\rm eff} = 0.999$ the maximum of
$C_{r,\phi,\eta}^{\hbox{\tiny (IPS)}}(J)$ is $2.40$.}
\end{figure}
The same conclusions holds when we consider the parameterization
of Eq. (\ref{bell:ale}). In Fig. \ref{f:ips:ale} we plot
$C_{r}^{\hbox{\tiny (TWB)}}(J)$ and $C_{r,\phi,\eta}^{\hbox{\tiny
(IPS)}}(J)$, {\em i.e.} $C(J)$ evaluated for the TWB and the IPS
state, respectively. The behavior is similar to that of $B(J)$,
the maximum violation being now $C_{r,\phi,\eta}^{\hbox{\tiny
(IPS)}}(J) = 2.40$ for $\tau_{\rm eff} = 0.999$ and $J = 1.6\,
10^{-4}$.
\par
Finally, notice that the maximum violation using IPS states is
achieved (for both parameterizations) when $\tau_{\rm eff}$
approaches unit and for values of $r$ smaller than for TWB.
\section{Nonlocality and homodyne detection}\label{s:homodyne}
The Wigner function $W_{r,\phi,\eta}(\alpha,\beta)$ given in
Eq. (\ref{ips:wigner}) is
not positive-definite and thus $\rho_{\hbox{\tiny IPS}}$
can be  used to test the violation
of Bell's inequalities by means of  homodyne detection, {\em i.e.}
measuring the quadratures $x_{\vartheta}$ and
$x_{\varphi}$ of the two IPS modes $a$ and $b$, respectively,
as proposed in Ref. \cite{sanchez}. In this case, if one discretizes the
measured quadratures assuming as outcome $+1$ when $x \ge 0$, and $-1$
otherwise, one obtains the following Bell parameter
\begin{eqnarray}\label{bell:homo}
S &=&
E(\vartheta_1,\varphi_1) + E(\vartheta_1,\varphi_2)\nonumber\\
&\mbox{}& \hspace{1cm}
+ E(\vartheta_2,\varphi_1) - E(\vartheta_2,\varphi_2)\,,
\end{eqnarray}
where $\vartheta_j$ and $\varphi_j$ are the phases of the two
homodyne measurements at the modes $a$ and $b$, respectively, and
\begin{equation}
E(\vartheta_j,\varphi_k) =
\int_{\mathbb{R}^2} d x_{\vartheta_j}\,d x_{\varphi_k}\,
{\rm sign}[x_{\vartheta_j}\, x_{\varphi_k}]\,
P(x_{\vartheta_j}, x_{\varphi_k})\,,
\end{equation}
$P(x_{\vartheta_j}, x_{\varphi_k})$ being the joint
probability of obtaining the two outcomes
$x_{\vartheta_j}$ and $x_{\varphi_k}$ \cite{sanchez}. As usual, violation
of Bell's inequality is achieved when $|S|>2$.
\par
\begin{figure}[t!]
\includegraphics[scale=.5]{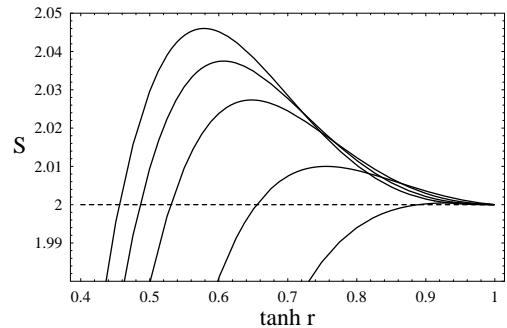}
\caption{\label{f:homo:id} Plots of $S$ given in Eq.
(\ref{bell:homo}) as a function of $\tanh (r)$ for different values of
$\tau_{\rm eff}$ and for ideal homodyne detection
({\em i.e.} with quantum efficiency $\eta_{\rm H}=1$): from top to bottom
$\tau_{\rm eff} = 0.99, 0.95, 0.90, 0.80$ and $0.70$.}
\end{figure}

In Fig. \ref{f:homo:id} we plot $S$ for $\vartheta_1 = 0$,
$\vartheta_2 = \pi/2$, $\varphi_1 = -\pi/4$ and $\varphi_2 = \pi/4$: as
pointed out in Ref. \cite{sanchez}, the Bell's inequality is violated
for a suitable choice of the squeezing parameter $r$. Notice that when
$\tau_{\rm eff}$ decreases the maximum of violation shifts toward higher
values of $r$.
\par
\begin{figure}[t!]
\includegraphics[scale=.5]{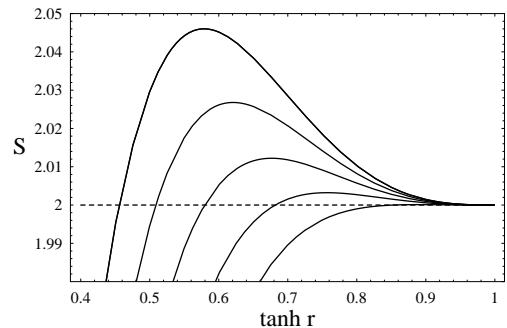}
\caption{\label{f:homo:eta} Plots of $S$ given in Eq.
(\ref{bell:homo}) as a function of $\tanh (r)$ with $\tau_{\rm eff} = 0.99$
and for different values of the  homodyne detection efficiency
$\eta_{\hbox{\tiny H}}$: from top to bottom
$\eta_{\hbox{\tiny H}} = 1, 0.95, 0.90, 0.85$ and $0.80$. The maximum of
the violation decreases and shifts toward higher values of $r$ as
$\eta_{\hbox{\tiny H}}$ decreases. For smaller values of $\tau_{\rm eff}$
the violation is furtherly reduced.}
\end{figure}

As one expects, taking into account the efficiency $\eta_{\hbox{\tiny H}}$
of the homodyne detection furtherly reduces the violation (see
Fig. \ref{f:homo:eta}). Notice that, when $\eta_{\hbox{\tiny H}}<1$, violation
occurs for higher values of $r$, although its maximum is actually
reduced: in order to have a significative violation one needs a
homodyne efficiency greater than 80\% (when $\tau_{\rm eff}=0.99$).
\section{Concluding remarks}\label{s:remarks}
In this paper we have shown that IPS can be used to produce non-Gaussian
two-mode states starting from a TWB. We have studied the nonlocality
of IPS states in phase-space using the Wigner function. As for the
improvement of IPS assisted teleportation \cite{ips:tele}, we have found
that the nonlocal correlations are enhanced for small energies of the
TWB (small squeezing parameter $r$). Moreover, nonlocality of
$\varrho_{\hbox{\tiny IPS}}$ is larger than that of TWB irrespective
of IPS quantum efficiency.
\par
Since the Wigner function of the IPS state is not positive
definite, we have also analyzed its nonlocality using homodyne
detection. In this case violation of Bell's inequality is much
less than in the phase-space, and is furtherly reduced for non
unit homodyne efficiency $\eta_{\hbox{\tiny H}}<1$. However, this
setup (IPS with homodyning) is of particular interest, since it
can be realized with current technology achieving a loophole-free
test of Bell's inequality \cite{sanchez}.
\par
On the other hand, the experimental verification of phase-space
nonlocality is challenging, due to the difficulties of measuring
the parity, either directly or through the measurement of the
photon distribution. On the other hand, the recent experimental
generation of IPS states \cite{grangier} is indeed a step toward
its implementation.
\section*{Acknowledgments}
SO would like to express his gratitude to A. R. Rossi and
A. Ferraro for stimulating discussions and for their continuous
assistance.

\end{document}